\documentclass{aa}  

\usepackage{graphicx}
\usepackage{xcolor}
\usepackage{txfonts}
\usepackage{amssymb}
\usepackage{graphicx}
\usepackage{amsmath}
\usepackage{pict2e}
\usepackage{multirow}
\usepackage{todonotes}

\newcommand{\Mc}[1]{#1}

\def\bK{\mbox{\boldmath$K$}}
\def\bu{\mbox{\boldmath$u$}}
\def\bxi{\mbox{\boldmath$\xi$}}
\def\br{\mbox{\boldmath$r$}}

\def\ii{{\rm i}}

\def\lb{\overline{\ell}}
\def\mb{\overline{m}}
\def\cl{\alpha_\ell}
\def\clp{\alpha_{\ell'}}
\def\clb{\alpha_{\lb}}

\begin{document} 

\title{Sensitivity kernels for time-distance helioseismology:  \\
efficient computation for spherically-symmetric solar models}
\titlerunning{Sensitivity kernels for time-distance helioseismology}

\author{
		Damien Fournier\inst{1}
        \and Chris S. Hanson\inst{1}
        \and Laurent Gizon\inst{1,2}
        \and H\'el\`ene Barucq\inst{3}
        }
          
\institute{
Max-Planck-Institut f\"ur Sonnensystemforschung, Justus-von-Liebig-Weg 3, 37077 G{\"o}ttingen,   Germany \\ \email{fournier@mps.mpg.de}    
\and
Institut f\"ur Astrophysik, Georg-August-Universit\"at G\"ottingen, Friedrich-Hund-Platz 1, 37077 G\"ottingen, Germany  
\and
Magique-3D, Inria Bordeaux Sud-Ouest, Universit\'e de Pau et des Pays de l’Adour, 64013 Pau, France
			}

   \date{Received \today; accepted  XXX}

% \abstract{}{}{}{}{} 
% 5 {} token are mandatory
 
  \abstract
   {The interpretation of helioseismic measurements, such as wave travel-time, is based on the computation of kernels that give the sensitivity of the measurements to localized changes in the solar interior. These are computed using the ray or the Born approximation. The Born approximation is preferable as it takes finite-wavelength effects into account, but can be computationally expensive.}
   {We propose a fast algorithm to compute travel-time sensitivity kernels under the assumption that the background solar medium is spherically symmetric.}
   {Kernels are typically expressed as products of Green's functions that depend upon depth, latitude and longitude. Here, we compute the spherical harmonic decomposition of the kernels and show that the integrals in latitude and longitude can be performed analytically. In particular, the integrals of the product of three associated Legendre polynomials can be computed  thanks to the algorithm of Dong and Lemus (2002).}
   {The computations are fast and accurate and only require the knowledge of the Green's function where the source is at the pole. The computation time is reduced by two orders of magnitude compared to other recent computational frameworks.}
   {This new method allows for flexible and computationally efficient calculations of a large number of kernels, required in addressing key helioseismic problems. For example, the computation of all the kernels required for meridional flow inversion takes less than two hours on 100 cores.}
   \keywords{Sun: helioseismology -- Sun: oscillations -- Sun: interior -- Methods: numerical               }

   \maketitle
%
%________________________________________________________________

%\tableofcontents

\section{Introduction}

Time-distance helioseismology \citep{DUV93} aims at inferring the subsurface structure of the Sun by measuring seismic wave travel times between any two points at the solar surface. The interpretation of these measurements requires understanding how waves propagate in the solar interior, i.e. solving the forward problem. Due to its simplicity, the ray approximation was initially used to invert for flow velocities and sound-speed perturbations to a reference background model \citep{KOS96}. It is still used nowadays, for example to recover the meridional circulation \citep{RAJ15}. However, this approach is a high-frequency approximation that cannot be used to recover perturbations with sizes of order of the local wavelength \citep{Birch2000}. %Moreover, it is not able to take into account the complex analysis performed with the observations (projections, filtering, averaging) which could lead to misinterpretation of the data.
\citet{GIZ02} derived a general framework for sensitivity kernels under the Born approximation and for random sources of excitation. \citet{Birch2004, Burston2015, BOE16} computed Born kernels using a normal-mode summation of the eigenfunctions in a solar-like stratified background. To treat axisymmetric background media (e.g., a background that includes large-scale differential rotation) and to include frequencies above the acoustic cut-off, \citet{GIZ17} proposed to solve the wave equation in frequency space using a 2.5D finite-element solver. All these approaches are useful but are computationally expensive, which limit their use in the interpretation of solar data as many kernels must be computed (and averaged). In some cases, it is sufficient to consider perturbations to a steady spherically-symmetric reference medium.
The study of meridional circulation is one such application \cite[see, e.g.,][]{LIA17}. 

In this paper, we present a way to reduce the computational time of  Born sensitivity kernels in a spherically-symmetric background by treating the horizontal variables (the co-latitude $\theta$ and the longitude $\phi$) analytically using the properties of the spherical harmonics. Here, this approach is demonstrated using the scalar wave equation from \citet{GIZ17}, but could be applied to the normal-mode summation method of \citet{BOE16}, or solving the wave equation using a high-order finite-difference scheme  \citep{MAN17}. %The theoretical derivation is given in Section~\ref{sect:derivation} and a comparison with the previous method underlying the computational benefits of this approach is presented in Section~\ref{sect:results}.

\section{Born sensitivity kernels} \label{sect:derivation}

\subsection{Green's function in a spherically symmetric background}

We follow the framework of \citet{GIZ17}, where the observable $\psi(\br,\omega)$ at spatial location $\br = (r,\theta,\phi)$ and frequency $\omega$ is linked to the divergence of the displacement: $\psi(\br,\omega) = c(\br) \nabla \cdot \bxi(\br,\omega)$. This scalar quantity solves
\begin{equation}
\Mc{L} \psi (\br,\omega) = s(\br,\omega), \label{eq:wave}
\end{equation}
where $L$ is the spatial wave operator at frequency $\omega$,
\begin{equation}
\Mc{L} \psi := -(\omega^2 + 2\ii\omega \gamma) \psi -2 \ii \omega \bu \cdot \nabla \psi -c \nabla \cdot \left( \frac{1}{\rho} \nabla (\rho c \psi) \right) ,
\end{equation}
$\rho$ and $c$ are the solar density and sound speed from standard solar model~S \citep{CHR96}, $\gamma$ is the attenuation, $\bu$ is a background flow, and $s$ is a stochastic source term. We assume that the sources are spatially uncorrelated and depend only on depth (and frequency) such that the source covariance matrix is given by
\begin{equation}
M(\br,\br',\omega) := \mathbb{E}[s^\ast(\br,\omega) s(\br',\omega)] =  A(r,\omega) \delta(\br-\br'), \label{eq:source}
\end{equation}
where $A(r,\omega)$ is the radial profile of the source power.
The wave field $\psi$ can be obtained using
\begin{equation}
\psi(\br,\omega) = \int_\odot G(\br,\br',\omega) s(\br',\omega) \rho(\br') \textrm{d}\br' ,
\end{equation}
where $G$ is the Green's function: \begin{equation}
\Mc{L}G(\br,\br',\omega) = \frac{1}{\rho(\br)} \delta(\br-\br') .
\end{equation}
When the background is spherically symmetric (i.e. no flow and no heterogeneity), $G$ can be written as
\begin{equation}
 G(\br, \br',\omega) = \sum_{\ell=0}^{\ell_{\rm max}} \cl  G_\ell(r, \br', \omega) \sum_{m=-\ell}^\ell Y_\ell^{m\ast}(\theta', \phi') Y_\ell^m(\theta, \phi), \label{eq:G}
 %&= \frac{1}{\sqrt{2\pi}} \sum_{l=0}^{L}  G_l(r; \br_0) P_l(\cos\gamma),
\end{equation}
where $\br=(r,\theta,\phi)$, $\br'=(r',\theta',\phi')$, $Y_\ell^m$ are the normalized spherical harmonics, $\cl = \sqrt{{4\pi}/(2\ell+1)}$, and $G_\ell$ is the Legendre component of the Green's function
\begin{equation}
G_\ell(r, \br', \omega) = \int_0^{2\pi} \int_0^\pi G(\br, \br',\omega) P_\ell(\cos\theta) \sin\theta \textrm{d}\theta \textrm{d}\phi. \label{eq:legendre}
\end{equation}
Equation~\eqref{eq:G} 
%with $\cos\gamma = \cos\theta \cos\theta_S + \sin\theta \sin\theta_S \cos(\phi+\phi_S)$.
could be simplified by using the addition theorem \citep[][Eq. 14.18.1]{NIST:DLMF} and introducing the great-circle angle between  $\br$ and $\br'$. However, Eq.~\eqref{eq:G} is the form required in the following sections.

\subsection{Cross-covariance in a spherically-symmetric background}

In a spherically-symmetric background the expectation value of the cross-covariance between an observation point $\br_1=(r_0,\theta_1,\phi_1)$ and a point $\br=(r,\theta,\phi)$     is 
\begin{align}
C(\br_1,\br,\omega) &= \mathbb{E}[\psi^*(\br_1,\omega) \psi(\br,\omega)] \nonumber \\ 
%=  \int G(\br_1, \br)^\ast G(\br_2,\br) f(r,\omega) d\br.  
&= \sum_\ell \cl^2 \sum_{m=-\ell}^\ell Y_\ell^{m\ast}(\theta_1,\phi_1) Y_\ell^m(\theta,\phi) C_\ell(\br_0,r,\omega), \label{eq:C}
\end{align}
where 
\begin{equation}
C_\ell(\br_0,r,\omega) = \int_0^{R_\odot} G_\ell(r',\br_0,\omega)^\ast G_\ell(r',\hat{\br},\omega)  A(r',\omega) \, \rho(r')^2 {r'}^2  \textrm{d}r' \label{eq:Cl}
\end{equation}
and $\br_0=(r_0,0,0)$ and $\hat{\br}=(r,0,0)$ are on the polar axis. The radius $r_0$ is the observation radius, for example $\sim$150~km above the photosphere for SDO/HMI. In obtaining Eq.~\eqref{eq:C}, we used the property that the cross-covariance depends only on the great-circle distance between the two points. To simplify the computations we place one point on the polar axis so that the Green's function is axisymmetric and only the mode $m=0$ needs to be computed.

Using the convenient source of excitation introduced in \citet{GIZ17}, $C_\ell(\br_0,r, \omega)$ is directly linked to the imaginary part of the Green's function $G_\ell(r,\br_0,\omega)$, but this assumption is not mandatory in this paper. The important assumption concerns the covariance of the sources of excitation that needs to be of the form given by Eq.~\eqref{eq:source}, such that the cross-covariance depends only on depths and the great-circle distance between the two points $\br_1$ and $\br$. This assumption is common in helioseismology and is generally used in forward modeling \citep[e.g.,][]{koso2000,BOE16,MAN17}. 

\subsection{Born flow kernels} \label{sect:kernel3D}

Recovering flows in the solar interior is a major goal for local helioseismology, hence we focus here on flow kernels. The method presented here could be applied to all other types of perturbations with respect to a spherically symmetric background. The Born sensitivity kernel $\bK = (K_{r}, K_{\theta}, K_{\phi})$ connects the travel-time perturbation $\delta\tau$ to the vector flow $\bu = (u_r,u_\theta, u_\phi)$, such that
\begin{equation}
\delta\tau (\br_1,\br_2) = \int_\odot \bK(\br,\br_1,\br_2)\cdot \bu(\br) \,  \text{d}\br .
\end{equation}
According to \cite{GIZ17} we have
\begin{align}\label{eq.flowKernel}
 \bK(\br,&\br_1,\br_2) =  2\ii \rho(r) \int_{-\infty}^\infty  \ \textrm{d}\omega  \ \omega  W^\ast(\br_1,\br_2,\omega) \times \nonumber \\
& \left[ G(\br_2,\br,\omega) \nabla C(\br_1, \br, \omega) - G^\ast(\br_1,\br,\omega) \nabla C^\ast(\br_2, \br, \omega) \right],
\end{align}
where $W$ is a weighting function that relates a change in the cross-covariance to a change in travel-time  \citep{GIZ02} and $\nabla = (\partial_r, 1/r \ \partial_\theta, 1/(r\sin\theta) \ \partial_\phi)$ is the gradient operator with respect to the scattering location $\br$. Note that in a spherically-symmetric background, seismic reciprocity implies $G(\br,\br',\omega)=G(\br',\br,\omega)$ for any $\br$ and $\br'$. The reference cross-covariance also satisfies $C(\br',\br,\omega)=C(\br,\br',\omega)$.

The expression for the kernel may differ when a different observable is chosen, however 
the above integral will always involve the product of a Green's function with the cross-covariance.
One approach \citep{BOE16,MAN17} to obtain the flow kernels is to compute the 3D Green's function and the cross-covariance using its spherical harmonic decomposition using Eq.~\eqref{eq:G}. 
A reference kernel is usually obtained for a fixed pair of observation points and later rotated to obtain kernels for other pairs of points.
%For a given separation distance between the points $\br_1$ and $\br_2$, it is possible to obtain the kernels for different central point by rotating the 3D kernel. 
However a fine resolution in $\theta$ and $\phi$ is required in order to perform this rotation accurately, which makes the computation expensive in both time and memory.

\subsection{Spherical harmonic decomposition of Born flow kernels}

In order to circumvent the disadvantages mentioned above (e.g. rotation) and improve accuracy, we propose a new approach based on the spherical harmonic decomposition of the kernel,
\begin{equation}
\bK(\br,\br_1,\br_2) = \sum_{\lb} \sum_{\mb=-\lb}^{\lb} \bK^{\lb \mb}(r,\br_1,\br_2) Y_{\lb}^{\mb}(\theta,\phi),
\end{equation}
where
\begin{equation}
\bK^{\lb \mb}(r, \br_1,\br_2) = \int_0^{2\pi} \int_0^\pi  \bK(\br, \br_1,\br_2) Y_{\lb}^{\mb \ast }(\theta,\phi) \sin\theta \textrm{d}\theta \textrm{d}\phi.
\end{equation}
Decomposing $G(\br_1,\br,\omega)$ and $C(\br_2,\br,\omega)$ into spherical harmonics, we can obtain the spherical harmonic coefficients of each kernel.

For the $u_r$ kernel, we have 
\begin{align}
& K_{r}^{\lb \mb}(r,\br_1,\br_2) = \sum_{\ell,\ell'}   \cl \clp \sum_{m=-\ell}^\ell \sum_{m'=-\ell'}^{\ell'} I_r \times \nonumber \\
&  \left( f^r_{\ell\ell'}(r) Y_\ell^{m*}(\theta_2,\phi_2) Y_{\ell'}^{m'*}(\theta_1,\phi_1) + g^r_{\ell\ell'}(r) Y_\ell^{m}(\theta_1,\phi_1) Y_{\ell'}^{m'}(\theta_2,\phi_2) \right), \label{eq:kernel1}
\end{align}
where
\begin{align}
f^r_{\ell\ell'}(r) &= 2 \ii \rho(r) \int_{-\infty}^\infty \ \omega 
W^\ast(\omega)G_\ell(r, \br_0, \omega) \partial_r C_{\ell'}(\br_0, r, \omega) \textrm{d}\omega, \\
g^r_{\ell\ell'}(r) &= -2 \ii \rho(r) \int_{-\infty}^\infty \  \omega
W^\ast(\omega)G_\ell^\ast(r, \br_0, \omega) \partial_r C_{\ell'}^\ast(\br_0, r, \omega) \textrm{d}\omega, \\
\text{and } I_r &= \int_0^{2\pi}\int_0^{\pi} Y_\ell^m(\theta,\phi) Y_{\ell'}^{m'}(\theta,\phi) Y_{\lb}^{\mb\ast}(\theta,\phi)\sin\theta \textrm{d}\theta \textrm{d}\phi. \label{eq:Ir}
\end{align}
The integral of three spherical harmonics over the unit sphere can be done analytically using the Gaunt formula \citep[see e.g.][ Eq.~(4.6.3)]{EDM60}
\begin{equation}
I_r= \frac{4\pi}{\cl \clp \clb} (-1)^{\mb} \left( \begin{array}{ccc} \ell & \ell' & \lb \\ 0 & 0 & 0 \end{array} \right) \left( \begin{array}{ccc} \ell & \ell' & \lb \\ m & m' & -\mb \end{array} \right), \label{eq:gaunt}
\end{equation}
where we have used the Wigner-3j symbols \citep[see e.g.][p.~45]{EDM60}. The Wigner-3j symbol vanishes when $\overline{m} \neq m+m'$, which enables us to remove the sum over $m'$ in  equation for $K_{u_r}$. 

It can be shown that the expression for $K_{r}$, $K_{\theta}$ and $K_{\phi}$ can be recast in the form
\begin{align}
 K_{j}^{\lb \mb}(r) =  \sum_{\ell,\ell'}   \cl \clp & \sum_{m=-L}^L   \Bigl( I_j f^j_{\ell\ell'}(r)  Y_\ell^{m*}(\theta_2,\phi_2) Y_{\ell'}^{\mb-m*}(\theta_1,\phi_1)  \nonumber \\
&  + I_j^\ast g^j_{\ell\ell'}(r)  Y_\ell^{m*}(\theta_1,\phi_1) Y_{\ell'}^{\mb-m*}(\theta_2,\phi_2) \Bigr) , \label{eq:Klm}
\end{align}
where $j\in\{r, \theta, \phi\}$ and $L = \min(\ell,\ell')$.
%\textcolor{red}{Check if we need the $*$ on the second Ylm Ylm.}

Proceeding in a similar way as for $K_{r}$, the kernel  $K_{\theta}^{\lb\mb}$ depends on the functions $f^\theta$ and $g^\theta$ given by
\begin{align}
f^\theta_{\ell\ell'}(r) &= - 2 \ii \rho(r) \int_{-\infty}^\infty \  \omega W^\ast(\omega)G_\ell(r, \br_0, \omega) C_{\ell'}(r,\br_0, \omega) \textrm{d}\omega, \\
g^\theta_{\ell\ell'}(r) &=  2\ii \rho(r) \int_{-\infty}^\infty \ \omega W^\ast(\omega)G_\ell^\ast(r, \br_0, \omega) C_{\ell'}^\ast(r,\br_0, \omega) \textrm{d}\omega.
\end{align}
The horizontal integral is 
\begin{equation}
I_\theta = \frac{1}{r} \int_0^{2\pi}\int_0^{\pi} Y_\ell^m(\theta,\phi) \partial_\theta Y_{\ell'}^{m'}(\theta,\phi) Y_{\lb}^{\mb *}(\theta,\phi) \sin\theta \textrm{d}\theta \textrm{d}\phi. \label{eq:Itheta}
\end{equation}
This integral $I_\theta$ is much more difficult to evaluate than $I_r$ because of the $\theta$ derivative.
%\CSH{Need to state that for $\ell=0$ the derivative is zero, otherwise we might get equation of why why we have $P^-1_0$}
In order to keep only associated Legendre polynomials in $I_\theta$, we use 
\begin{align}
\frac{d \Mc{P}_\ell^m(\cos\theta)}{d\theta} = \frac{1}{2}  & \left(  \sqrt{(\ell+m) (\ell-m+1)}  \Mc{P}_{\ell}^{m-1}(\cos\theta) \right. \nonumber \\
&  - \left. \sqrt{(\ell+m+1)(\ell-m)} \Mc{P}_\ell^{m+1}(\cos\theta) \right), \label{eq:dthetaPlm}
\end{align}
%\CSH{I think the $\sin\theta$ should not be in front. Once you transform the recurance relation to be in terms of $\theta$ not $x$ the $\sin\theta$ disappears}
where the $\Mc{P}_l^m$ are the normalized associated Legendre polynomials. We use the convention that $\Mc{P}_\ell^{m \pm 1} =0$ if $|m \pm 1| > \ell$, so that Eq.~\eqref{eq:dthetaPlm} remains valid for $m=\pm \ell$. %, $\int_{-1}^1  \Mc{P}_l^m(x)^2 \text{d}x = 1$.
Integrating Eq.~\eqref{eq:Itheta} over $\phi$ and using Eq.~\eqref{eq:dthetaPlm}, $I_\theta$ becomes
\begin{align}
I_\theta = \frac{1}{2\sqrt{2\pi} \ r} \Bigl( &-\sqrt{(\ell'+m')(\ell'-m'+1)} J_{\ell\ell'\lb}^{m,m'-1,\mb} \nonumber \\ 
& + \sqrt{(\ell'+m'+1)(\ell'-m')} J_{\ell\ell'\lb}^{m,m'+1,\mb} \Bigr), 
\end{align}
where $m'=\mb-m$ and 
\begin{equation}
J_{\ell\ell'\lb}^{mm'\mb} = \int_0^\pi \Mc{P}_\ell^m(\cos\theta) \Mc{P}_{\ell'}^{m'}(\cos\theta) \Mc{P}_{\lb}^{\mb}(\cos\theta)\sin\theta \textrm{d}\theta.
\end{equation}
Fortunately, this integral can also be evaluated analytically. It involves a sum of products of Wigner-3j symbols \citep[see Appendix~\ref{sect:appendix} and][]{DON02}.

%We note that the integral $I_\theta$ is more difficult to evaluate than $I_r$, since the sum of the three azimuthal orders, $m+m'+\mb$, is non-zero. 
%However, we will show in the next section that the evaluation of $I_\theta$ can be done efficiently.

The derivation of $u_\phi$ is similar to $u_\theta$ and requires the evaluation of the horizontal integral
\begin{equation}
I_\phi = \frac{\ii m'}{r} \int_0^{2\pi}\int_0^{\pi} \frac{1}{\sin\theta} Y_\ell^m(\theta,\phi) Y_{\ell'}^{m'}(\theta,\phi) Y_{\lb}^{\mb *}(\theta,\phi) \sin\theta \textrm{d}\theta \textrm{d}\phi.
\end{equation}
%\CSH{Need to explicitly state that for $m=0$ the derivative is zero, otherwise we might get equation of what happens for $1/m$}
%A $\sin\theta$ term is required in order to be able to evaluate this integral analytically. 
Using
\begin{align}
\frac{\Mc{P}_\ell^{m}(\cos\theta)}{\sin\theta}  =& -\frac{1}{m}   \left(  \sqrt{\frac{(2\ell+1)(\ell+m+1)(\ell+m)}{2\ell-1}}  \Mc{P}_{\ell-1}^{m-1}(\cos\theta) \right. \nonumber \\
& + \left. \sqrt{\frac{(2\ell+1)(\ell-m+1)(\ell-m)}{2\ell-1}} \Mc{P}_{\ell-1}^{m+1}(\cos\theta) \right), \label{eq:dphiPlm}
\end{align}
for $m \neq 0$, we obtain
\begin{align}
I_\phi = \frac{\ii}{2\sqrt{2\pi} \ r}  &\left( \sqrt{\frac{(2\ell+1)(\ell+m+1)(\ell+m)}{2\ell-1}} J_{\ell,\ell'-1,\lb}^{m,\mb-m-1,\mb} \right. \nonumber \\ 
& + \left. \sqrt{\frac{(2\ell+1)(\ell-m+1)(\ell-m)}{2\ell-1}} J_{\ell,\ell'-1,\lb}^{m,\mb-m+1,\mb} \right). \label{eq:Iphi}
\end{align}

%The integral for $I_r$ can be written explicitly  using the Gaunt formula, while the integrals for $I_\theta$ and $I_\phi$ require the evaluation of several Wigner-3j symbols using the algorithm presented in \citep{DON02} and given by Eq.~\eqref{eq:3Plm}. 

\begin{table*}[!htb]
\caption{Terms required to compute the spherical harmonic coefficients of the flow sensitivity kernels $K_{j}^{\lb\mb}$ using Eq.~\eqref{eq:Klm}. The integrals $J_{\ell\ell'\lb}^{mm'\mb}$ depend only on Wigner-3j symbols and can be computed using Eq.~\eqref{eq:3Plm} corresponding to the algorithm of \citet{DON02}.} \label{tab:summary}
\centering
\begin{tabular}{ccc} 
\hline\hline
$j$
&  $f^j_{\ell\ell'}(r)$ &  $I_j$ \\
\hline
$r$ & $ -2\ii\rho(r)\int_{-\infty}^\infty \  \omega W^\ast(\omega)G_\ell(r, \br_0, \omega) \partial_r C_{\ell'}(\br_0,r, \omega) d\omega$ 
& 
$\frac{4\pi}{\cl \clp \clb} \left( \begin{array}{ccc} \ell & \ell' & \lb \\ 0 & 0 & 0 \end{array} \right) \left( \begin{array}{ccc} \ell & \ell' & \lb \\ m & \mb-m & -\mb \end{array} \right)$ \\
\\[0.2em]
\multirow{2}{*}{$\theta$} & \multirow{2}{*}{$ -2\ii\rho(r)\int_{-\infty}^\infty \ \omega W^\ast(\omega)G_\ell(r, \br_0, \omega) C_{\ell'}(\br_0,r, \omega) d\omega$ }
&
$\frac{1}{2\sqrt{2\pi}r} \Bigl( -\sqrt{(\ell'+\mb-m)(\ell'-\mb+m+1)}  \ J_{\ell\ell'\lb}^{m,\mb-m-1,\mb} $ \\
& & \hspace*{1cm} $+\sqrt{(\ell'+\mb-m+1)(\ell'-\mb+m)} \ J_{\ell\ell'\lb}^{m,\mb-m+1,\mb} \Bigr)$ \\
\\[0.2em]
$\phi$ & $f^\phi_{\ell\ell'}(r) = f^\theta_{\ell\ell'}(r)$ &
$ \frac{\ii}{2\sqrt{2\pi}r}\Bigl( \sqrt{\frac{(2\ell'+1)(\ell'-\mb+m-1)(\ell'-\mb + m)}{2\ell'-1}} \ J_{\ell,\ell'-1,\lb}^{m,\mb-m+1,\mb}$ \\ & & \hspace*{1.5cm} 
$ + \sqrt{\frac{(2\ell'+1)(\ell'+\mb-m-1)(\ell'+\mb-m)}{2\ell'-1}} \ J_{\ell,\ell'-1,\lb}^{m,\mb-m-1,\mb}\Bigr)$\\
\hline
\end{tabular}
\end{table*}

Now that we have the equations for the kernels, let us summarize the algorithm used for the resolution:
\begin{enumerate}
\item Computation and storage of the Green's function $G_l(r, \br_0, \omega)$ with the source on the polar axis, as a function of depth and harmonic degree $\ell$ for all frequencies.
\item For each great-circle distance between $\br_1$ and $\br_2$: 
\begin{itemize}
\item computation of the cross-covariance using Eq.~\eqref{eq:C}. If one wants to use a convenient source of excitation of \citet{GIZ17}, the cross-covariance is directly obtained from the imaginary part of the Green's function.
\item computation of the weighting function $W$.
\item computation of the functions $f^j$ and $g^j$. 
\end{itemize}
\item Evaluation of the integrals $I_j$ and computation of the kernel using Eq.~\eqref{eq:Klm}. 
\end{enumerate}

A summary of the different terms required to compute the different components of the flow kernels using Eq.~\eqref{eq:Klm} is given in Table~\ref{tab:summary}.

%The evaluation of the Green's function in the first step can be replaced by the use of the eigenfunctions if one wants to use the method presented in \cite{BOE16} for example.

We note that the algorithm presented above could of course be used to compute sensitivity kernels for the cross-covariance amplitude using the linear definition of \citet{NAG17} and the appropriate choice of $W$. One can also get kernels for the cross-covariance function at a given  frequency by removing the weighting function. In this case, the functions $g^j$ are just the complex conjugates of the $f^j$.

\begin{figure}
\includegraphics[width=\linewidth]{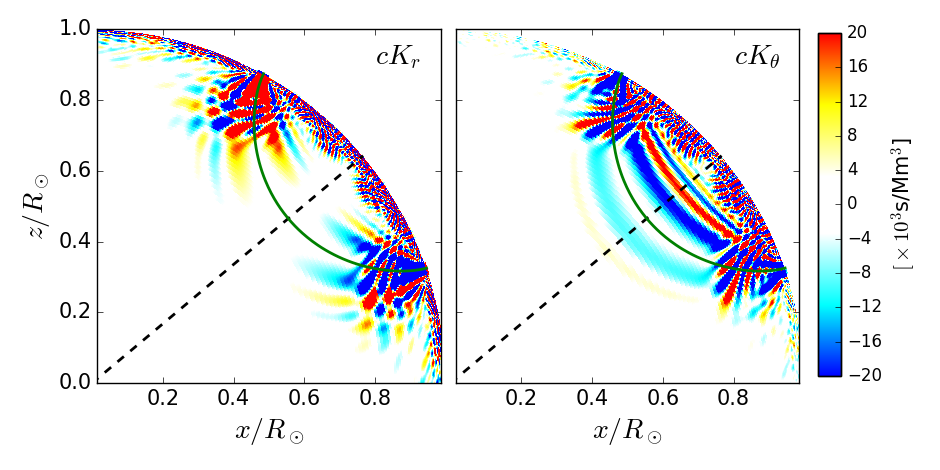}
\caption{Slices along a constant meridian of the point-to-point 3D travel-time difference kernel for $u_r$ (left) and $u_\theta$ (right). The 3D kernel for $u_\phi$ is zero along this slice. The kernel is computed with $\br_1$ and $\br_2$ separated by $42^\circ$, with mean latitude  $40^\circ$. The green line is the ray path between the two points and the dashed black line shows the image plane of Fig.~\ref{fig.3dKernel_S}.}
\label{fig.3dKernel_Y}
\end{figure}

\begin{figure}
\includegraphics[width=\linewidth]{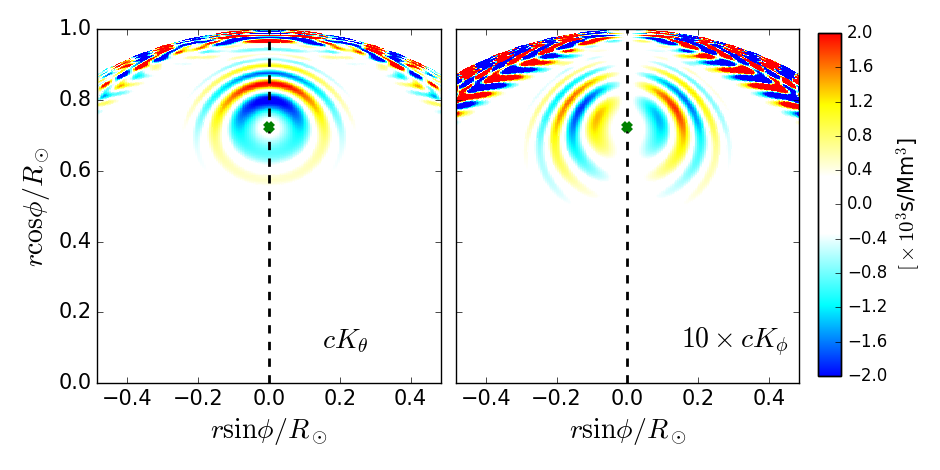}
\caption{Slices of the point-to-point 3D travel-time difference kernel for $u_\theta$ (left) and $u_\phi$ (right) along the plane indicated in Fig.~\ref{fig.3dKernel_S}. The 3D kernel for $u_r$ is mostly zero within this plane. The kernel is computed with $\br_1$ and $\br_2$ separated by $42^\circ$, with mean latitude  $40^\circ$. The green cross indicates the intersection of the ray path and the image plane. The dashed black line shows the image plane of Fig.~\ref{fig.3dKernel_Y}.}
\label{fig.3dKernel_S}
%\textcolor{red}{The choice of coordinate, $\phi$, in these panels is no good. Use $\chi$?}
\end{figure}

\subsection{Numerical validation}

To evaluate the flow kernels in this framework, the only ingredient to prescribe is the Green's function as a function of the spherical harmonic degree $\ell$ and depth for a source located at the pole. We compute it using 1D finite elements with the solver Montjoie \citep{CHA16,FOU17}. The Green's functions are computed with a high enough frequency resolution to resolve the modes using the mode linewidths  (5671 frequencies corresponding to 4 days of observations at 60~s cadence) and $\ell_{\rm max} = 400$.

A representation of the different components of the flow kernels between two points $\br_1$ and $\br_2$ centered at $40^\circ$ and separated by $42^\circ$ is shown in Figs.~\ref{fig.3dKernel_Y}~and~\ref{fig.3dKernel_S}. They exhibit the classical banana-doughnut shape with zero sensitivity along the ray path. Small scale structures are visible close to the surface as we kept values of $\ell$ up to 400. Visually, there is no difference between the kernels computed with this new approach, the ones from \citet{GIZ17} or the ones obtained by rotation so only one is shown here.  

\begin{figure}
% \missingfigure[figwidth=\linewidth]{Testing a long text string}
\includegraphics[width=\linewidth]{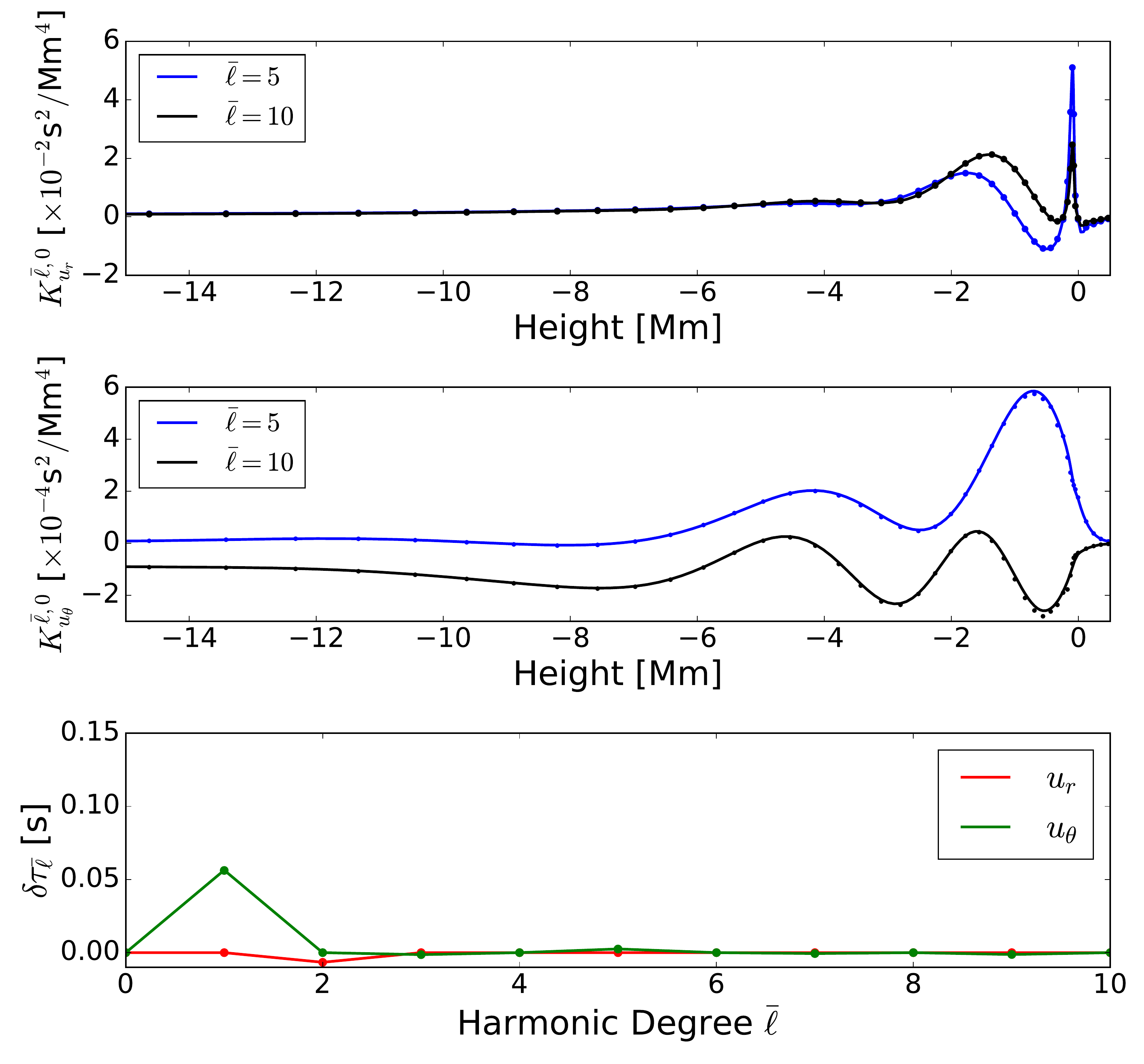}
\caption{Top and middle panels: Comparison of the kernels for $\bar{\ell} = 5$ and 10 (blue and black, respectively) computed using the method here (solid lines) and the method of \citet{GIZ17} (dots). Bottom panel: The travel times $\delta \tau_{\lb}$ due to the radial (red) and the latitudinal (green) components of the flow for each $\bar{\ell}$ of the kernels presented here (solid line) and those of \citet{GIZ17} (dots).}
\label{fig.compareKell}
\end{figure}

To allow a more quantitative approach, Figure~\ref{fig.compareKell} compares kernels computed using our new approach to the one presented by \citet{GIZ17} where the background is axisymmetric and the Green's function is computed for each azimuthal degree $m$ on a 2D grid. These kernels $K_r$ and $K_\theta$ are averaged over longitudes ($\bar{m} = 0$) where $\br_1$ is located at the pole and $\br_2$ is at a co-latitude of $42^\circ$ \citep[akin to Fig.~(17) of ][]{GIZ17}.  The results show good agreement, validating the method presented here. We note  slight differences in the structure of  $K_\theta$ at a depth of 500~km, and attribute this to the numerics of the 2D FEM solver differing. Specifically, the 2D FEM has inherent difficulty to compute the real part of the Green's function close to the Dirac source location \citep[see][for details]{CHA16}. In order to ensure that these small differences are not affecting the interpretation of the data, we compute the travel times induced by the meridional flow model from \citet{GIZ17}. We decompose the flow in Legendre polynomial $\bu^{\lb}$ akin to Eq.~\eqref{eq:legendre} and compute a travel time for each $\lb$ according to
\begin{equation}
\delta \tau_{\lb} = \int_0^{R_\odot} \bK^{\lb,\mb=0}(r) \cdot \bu^{\lb}(r) \, r^2 \textrm{d}r .
\end{equation}
The bottom panel of Fig.~\ref{fig.compareKell} shows this travel time as a function of $\lb$ is nearly indistinguishable from the travel times from \citet{GIZ17}, with differences less than $0.5$~ms. 

%------------------------------------------------------------------------
\section{Computational cost} \label{sect:results}

As the computational burden of this new method depends on the maximal harmonic degree of the Green's function $\ell_{\rm max}$ and the number of $\lb$ required for the kernel, we illustrate the efficiency of the method on two problems of interest: kernels for  meridional flow inversions and kernels for supergranulation inversions.

\subsection{Computation of kernels for meridional flow}

As a first test, we compute the kernels that are required to interpret meridional flow measurements. As the flow varies slowly with latitude, we can limit the number of spherical harmonic coefficients of the kernels to $\lb \le 10$ (see Fig.~\ref{fig.compareKell}). As for the observations, a low-pass filter with $\ell_{\rm max}=300$ is applied to the Green's function. The method can be parallelized in $\lb$ so we use 11 cores to compute kernels up to $\lb=10$.

For meridional flow measurements, one generally prescribes the separation distance between the source and the receiver for different values of the mean latitude \citep[see e.g.][]{LIA17}. For a given separation distance, we compute 15 kernels corresponding to 15 different latitudes. We then vary the separation distance in order to probe different depths.

Table~\ref{tab:steps} shows the computational times and memory requirements of the different steps of the algorithm. The computation of the Green's function for a source located at the pole can be done once and for all and stored as it is necessary to do it for every kernel. Therefore the computation is very fast  (3 seconds per frequency for $\ell_{\rm max} = 300$) and embarassingly parallel in frequency so it could also be recomputed every time. The computation of the frequency integrals ($f^j$ and $g^j$) consists in loading the Green's function and the weighting function $W$ and summing over frequencies. Major part of the time is due to reading  the Green's function files for all frequencies. The spatial integrals $I_r$, $I_\theta$ and $I_\phi$ could be computed once for all and stored for future use. However the computational time is small compared to the full computation of the kernel so we decide to recompute $I_j$ every time as the reading time can depend upon file system I/O load. The computations are parallelized in $\lb$ and hence need 11~cores for each step since $0\leq \lb \leq 10$. The computation of the kernel for $u_r$ is faster since the computation of $I_r$ requires the evaluation of only two Wigner-3j symbols, unlike $I_\theta$. However the major difference in computational time between $K_r$ and $K_\theta$ comes from the sum in $\ell'$ in Eq.~\eqref{eq:Klm}. For $K_r$, the sum covers the range from $\ell-\lb$ to $\ell+\lb$, since $I_r=0$ for other values due to the properties of the Wigner-3j. On the contrary, the sum in $\ell'$ for the computation of $K_\theta$ must be computed for the full range from 0 to $\ell$. 

\begin{table}
\caption{Computational time of the different steps to obtain 15 flow kernels (15 latitudes) for a given separation distance with $\lb \leq 10$ and $\ell_{\rm max}=300$ using 11 cores.}
\label{tab:steps}
\centering
\begin{tabular}{lcc}
\hline\hline 
{Computation steps} & Time [11 cores] & Memory \\
\hline
Green's function & 13~min & 100~MB \\
 $f^r$ (or $f^\theta$) & 9~min & 1~GB \\
 $I_r$ & 0.5~s & 1~GB \\
$I_\theta$ & 6~min & 1~GB \\
Sum of terms in $K^{\lb\mb}_{r}$  & 3~min & 1~GB\\
Sum of terms in $K^{\lb\mb}_{\theta}$  & 32~min & 1~GB \\
\hline
\end{tabular}
\end{table}

Even though the computational burden of $K_{\theta}$ is greater than $K_{r}$, the total burden remains significantly lower than for other methods, see Table~\ref{tab:comparison}. All the kernels required to perform a meridional flow inversion can be computed within 2~hours with 100 cores and the memory requirements do not exceed 1~GB. The approach mentioned  in Sect.~\ref{sect:kernel3D}, where the full 3D kernel is computed and rotated to obtain different latitudes, would take 11 days using 1000 cores with very significant memory requirements. In the axisymmetric approach of \citet{GIZ17} the computational time %depends on the separation distance between to the observation points as fewer azimuthal orders are required to converge at large separation distances. 
%For example, only 30~m are required for a separation distance of $42^\circ$ while 200 are necessary for $6^\circ$ separation. 
would be about 40 days on 1000 cores for all the same set of kernels.

\begin{table}
\caption{Comparison of the computational time and memory requirements to evaluate 225 kernels for the meridional flow using the method presented in this paper, the rotation of the 3D kernels as in \citet{BOE16, MAN17} using a horizontal grid sampled with $N_\theta=1001$ and $N_\phi=2001$ points, and the approach of \citet{GIZ17}.}
\label{tab:comparison}
\centering
\begin{tabular}{lcc} \hline \hline
Method & Time [cpu hours] & Memory [GB] \\
\hline
This paper & 170 & 1 \\
Rotation of 3D kernels &2.7$\times10^5$ & 40 \\
\citet{GIZ17} & $10^6$ & 8 \\ \hline
\end{tabular}
\end{table}

The computational times presented here are for point-to-point measurements, however, this framework can easily be extended to geometric averaging such as arc-to-arc measurements often performed for meridional flow measurements \citep[e.g,][]{LIA17}. One only needs to replace the product of the two spherical harmonics in Eq.~\eqref{eq:Klm} by a sum over all the points of the arc.

\subsection{Computation of kernels for supergranulation}

Resolving smaller scale flows, such as supergranules, requires a high spatial resolution and thus the Green's function needs to include much higher harmonic degrees than for meridional flow Green's function. For example, \citet{DUV13} considered measurements up to $\ell_{\rm max}=700$, but even higher $\ell_{\rm max}$ values may be required. Furthermore supergranulation flows have maximum power  around $\ell=120$, thus the kernels should at least be computed up to $\lb = 300$, or higher depending on the power distribution of the flow at large $\ell$. The computational burden for these kernels is summarized in Table~\ref{tab:stepsSupergranule}. The computation of the $m=0$ component of the Green's function now takes about 23~s per frequency, and the loading of the files to compute $f^j$ around 6~min. The computation of the Wigner symbols is computationally more challenging as $\ell$ increases, since the number of loops scales as $\ell_{\rm max}^3$ due to loops in $\ell$, $\ell'$ and $m$. While the computation of $I_r$ is still fast, the evaluation of $I_\theta$ now takes 270~min on 100 cores. 
%The biggest increase in the kernel computation time is due to the computation of the kernel itself which now takes 15~h for $\bu_r$ and 45~h for $bu_\theta$. 
Computing a set of 200~kernels would take 3~days with 1000~cores which is significantly longer than for the meridional flow kernels, but still one order of magnitude faster than the approach of \citet{GIZ17} and with a smaller memory requirement. 

\begin{table}
\caption{Computational time of the different steps to obtain 15 flow kernels for a given separation distance with $\lb \leq 300$ and $\ell_{\rm max}=700$ using 100 cores.}
\label{tab:stepsSupergranule}
\centering
\begin{tabular}{lcc} \hline \hline
{Computation steps} & Time [100 cores] & {Memory} \\
\hline
Green's function & 11~min & 100~MB \\
 $f^r$ (or $f^\theta$) & 6~min & 1~GB \\
 $I_r$ & 2~min & 1~GB \\
 $I_\theta$ & 270~min & 1~GB \\
Sum of terms in $K^{\lb,\mb}_{r}$
& 15~h & 1~GB \\
Sum of terms in $K^{\lb,\mb}_{\theta}$ &39~h & 1~GB \\ \hline
\end{tabular}
\end{table}

\section{Conclusions}

%The poor signal-to-noise of helioseismic travel times has led observers to perform various spatial averages, e.g., arc-to-arc, center-to-annulus, or deep focusing. This leads to the computation of large numbers of sensitivity kernels. 

We presented a technique faster than previous approaches to compute travel-time kernels under the assumption that the background medium is spherically symmetric. This technique does not rely on the numerical computation of kernel rotations and thus does not require large memory. Instead the spatial integrals are performed analytically, which also leads to higher accuracy. For example, for meridional circulation applications, the kernels can be computed one thousand times faster than with previous methods, using a tenth of the memory requirement.

%Here we have shown that our technique dramatically reduces computational burden, with the required CPU hours being just one hundredth of those demanded by existing methods. Additionally, the computation of complex geometry averaging kernels adds little to the computing time since the averaging is done in a precomputed factor governing the positions of the sources and receivers on the arc. This new method  will help to efficiently produce kernels that are related to the various averaging schemes of the observations, thus further improving the reliability of forward models.

\begin{acknowledgements}
The computer infrastructure was provided  by the German Data Center for SDO funded by the German Aerospace Center (DLR) and by the Ministry of Science of the State of Lower Saxony, Germany.
%The authors acknowledge financial support for computing resources from the Ministry of Science of the State of Lower Saxony, Germany.
\end{acknowledgements}

\begin{appendix}

\section{Algorithm for the integral of three associated Legendre polynomials} \label{sect:appendix}

For the sake of completeness, we summarize the algorithm of \citet{DON02} adapted to this study. The integral of three associated Legendre polynomials,
\begin{equation}
J_{\ell\ell'\lb}^{mm'\mb} = \int_0^\pi \Mc{P}_\ell^m(\cos\theta) \Mc{P}_{\ell'}^{m'}(\cos\theta) \Mc{P}_{\lb}^{\mb}(\cos\theta)\sin\theta d\theta ,
\end{equation}
can be computed analytically in terms of sums of products of Wigner-3j symbols:
\begin{align}
J_{\ell\ell'\lb}^{mm'\mb} &= \frac{(-1)^{\mb} (2\pi)^{3/2}}{\alpha_{\ell} \alpha_{\ell'} \alpha_{\lb} } \sum_{\ell_{12}=\min(|\ell-\ell'|, m_{12})}^{\ell+\ell'} Q_{12} \ \times \nonumber \\ 
& \sum_{\ell_{123}=\min(|\ell_{12}-\lb|, m_{123})}^{\ell_{12}+\lb}  Q_{123} \sqrt{\frac{(\ell_{123}-m_{123})!}{(\ell_{123}+m_{123})!}} J(\ell_{123},m_{123}), \label{eq:3Plm}
\end{align}
where the indices $m_{12} = m+m'$ and $m_{123}=m+m'+\mb$ represent sums over the azimuthal degrees. The quantities $Q_{12}$ and $Q_{123}$ must be evaluated for various values of $\ell_{12}$ and $\ell_{123}$ as defined under the sums in Eq.~\eqref{eq:3Plm}. They depend on the Wigner-3j symbols:
\begin{align*}
Q_{12} &=  (2\ell_{12} + 1) \left( \begin{array}{ccc} \ell & \ell' & \ell_{12} \\ 0 & 0 & 0 \end{array} \right) \left( \begin{array}{ccc} \ell & \ell' & \ell_{12} \\ m & m' & -m_{12} \end{array} \right), \\
Q_{123} &=  (2\ell_{123} + 1) \left( \begin{array}{ccc} \ell_{12} & \lb & \ell_{123} \\ 0 & 0 & 0 \end{array} \right) \left( \begin{array}{ccc} \ell_{12} & \lb & \ell_{123} \\ m_{12} & \mb & -m_{123} \end{array} \right).
\end{align*}
$Q_{12}$ (resp. $Q_{123}$) is non-zero only if $\ell_{12}+\ell+\ell'$ (resp. $\ell_{12}+\lb+\ell_{123}$) is even. 
%\textcolor{red}{Redefine $\ell_{12}$ and $\ell_{123}$ in text.}
The last term $J(\ell_{123},m_{123})$ is the integral  
\begin{equation}
J(\ell_{123},m_{123}) = \int_{-1}^1 \Mc{P}_{\ell_{123}}^{m_{123}}(x) \textrm{d}x ,
\end{equation}
which can be evaluated analytically. In this paper, we only need this value for $m_{123} = \pm 1$. As this integral is zero for odd values of $\ell_{123}$ due to the parity of the associated Legendre polynomials, we set $\ell_{123} = 2p+1$. Then, for a given $m_{123}$, the value of $J(\ell_{123},m_{123})$ can be evaluated recursively using
\begin{eqnarray}
J(2n+1,1) &=& \frac{(2n+1)(2n-1)}{4n(n+1)} J(2n-1,1)  \\
\text{and  }\; J(2n+1,-1) &=& \frac{(2n-1)^2}{4(n+1)^2} J(2n-1,-1), 
\end{eqnarray}
where $n=1, 2, \cdots p$, together with the initial conditions 
\begin{equation}
 J(1,1) = -\frac{\pi}{2} \; \text{ and  }\; 
J(1,-1) = \frac{\pi}{4}. 
\end{equation}
\end{appendix}

\bibliographystyle{aa}
\bibliography{biblio}

\end{document}